\documentclass[amsmath,amssymb,letterpaper,11pt]{article}
\usepackage{epsfig}
\usepackage{graphicx}
\usepackage{rotating}
\usepackage{amssymb}
\usepackage{amsmath}
\usepackage{subfigure}
\usepackage{titling, authblk}
\usepackage{fullpage}

\newtheorem{theorem}{Theorem}[section]

\newtheorem{proposition}[theorem]{Proposition}

\newtheorem{corollary}[theorem]{Corollary}

\newtheorem{assumption}[theorem]{Assumption}

\newcommand{\Level}{\mathit{lvl}}
\newcommand{\Aggregate}{\mathit{agg}}
\newcommand{\Parent}{\mathit{par}}

\newcommand{\Nat}{\omega}
\newcommand{\Neighbours}{n}
\newcommand{\Nxus}{N^{\mathit{us}}_x}
\newcommand{\Nus}{N^{\mathit{us}}}
\newcommand{\Ns}{N^s}

\begin{document}

\title{The Accuracy of Tree-based Counting in Dynamic Networks}
\author[1,2]{Supriya Krishnamurthy}
\author[1]{John Ardelius}
\affil[1]{SICS Center for Networked Systems (CNS), Stockholm, Sweden}
\author[2,3]{Erik Aurell}
\author[2]{Mads Dam}
\author[2]{Rolf Stadler}
\author[2]{Fetahi Wuhib}
\affil[2]{ACCESS Linnaeus Center, KTH - Royal Institute of Technology, Stockholm, Sweden}
\affil[3]{Dept. Information and Computer Science, Aalto University, Espoo, Finland}

\date{}
\maketitle

\begin{abstract}
Tree-based protocols are ubiquitous in distributed systems. They are
flexible,
they perform generally well, and, in static conditions, their analysis is
mostly simple. Under churn, however, node joins and failures can have
complex
global effects on the tree overlays, making analysis surprisingly subtle.
To our knowledge, few prior analytic results for performance estimation of
tree
based protocols under churn are currently known. We study a simple
Bellman-Ford-like protocol which performs network size estimation over a
tree-shaped overlay. A continuous time Markov model is constructed which
allows key protocol characteristics to be estimated, including the
expected
number of nodes at a given (perceived) distance to the root and, for each
such
node, the expected (perceived) size of the subnetwork rooted at that
node. We
validate the model by simulation, using a range of network sizes, node
degrees, and churn-to-protocol rates, with convincing results.
\end{abstract}

\thispagestyle{empty}
%\setcounter{page}{0}
%\pagebreak
%\part*{}

%\title{The Accuracy of Tree-based Counting in Dynamic Networks}

%\date{}
%\maketitle

\section{Introduction}

\label{sec:intro}

Tree-based protocols are used for many purposes in distributed
computing and networking, including routing, multicasting, distributed
monitoring, fault management, and data aggregation. Protocols that construct
and maintain tree overlays for these purposes are often simple, and they are
easy to analyze as long as the underlying network (graph) is static. In
the case of node churn, however, tree maintenance can cause highly complex
behavior. In fact, at sufficiently high churn rates, the overlays will never
stabilize to a proper spanning tree, and the underlying network may become
disconnected. In the case of an aggregation protocol, these phenomena can
cause significant errors when estimating the system size, for instance. In
this paper, our objective is to estimate this error, and, more generally, to
present an analytic model of the tree maintenance and aggregation processes
using continuous-time Markov models.

\paragraph{The Protocol.}
We study the problem in the context of a simple, tree-based aggregation
protocol named GAP \cite{GAP}, which is an asynchronous adaptation and
extension of the self-stabilizing BFS construction algorithm of Dolev et al
\cite{dolev1993self}. GAP is self-stabilizing in the sense that, once no
churn occurs, the network and aggregation data eventually become static, the
overlay forms a stable BFS tree, and the root has the correct aggregate, in
time linear in the diameter of the network graph. A tree based aggregation
protocol allows computing aggregates of local variables, such as node
counts, max/min, sums, averages, distributions, and distributed selection
queries, such as top $k$ or $k$'th smallest values, with good efficiency
(cf. \cite{DBLP:journals/cacm/KuhnLW08}). In this paper we restrict the
discussion to counting, and the protocol thus estimates the number of nodes,
or the network size. 

\paragraph{The Model.}
We model the protocol behavior using the master-equation approach of
statistical physics, cf. \cite{vanKampen}. For tractability we assume
a standard Poisson-arrival model with equal join and failure rates
$\lambda_j$ and $\lambda_f$, in order to achieve a stochastic equilibrium.
Each node maintains a local aggregate, a level variable indicating its
belief of the distance to the root, and a pointer to its current parent. At
a Poissonian rate $\lambda_g$, a node performs a protocol cycle, which
includes sampling the state of its neighbors and updating its own state
variables accordingly. We construct a model to estimate the expected value
of the local aggregate for a node at level $x$ ($x$ indicating the distance
in hops of the node from the root). The model is built in two steps. We
first estimate the expected values of the quantities $N_x$, the number of
nodes at level $x$, which we then use to estimate the quantities $M_x$, the
expected values of the local aggregate of nodes at level $x$. Given that
$N_0 = 1$ by construction, $M_0$ represents the estimate of the global
aggregate at the root, or, in the case of our protocol, the estimate of the
network size.

\paragraph{The Evaluation.}
The model is evaluated through simulation, for network sizes ranging from
$10^3$ to $10^5$ and rates $r=\lambda_g/\lambda_f$ between $1$ and $1,000$.
In the case of $r=1$ the tree overlay is highly unstable, and the protocol
behavior is dominated by the join subprotocol. For $r=1,000$,
tree-maintenance is performed at a much higher rate, and a node executes an
average of $1,000$ protocol cycles during its lifetime. Clearly, the effect
of churn in the system becomes more visible the lower the value of $r$
becomes\footnote{Although, perhaps surprisingly, the effect of
disconnection on level assignments is smaller.}.
The simulations show a good fit of the measurements to the analytic model
for the parameter ranges examined.

\paragraph{Related Work.}
A number of authors have studied analytical models of churn processes and
their effects on the stability and performance of p2p systems, cf.
\cite{Gummadietal03,StutzbachR06,LibenNowellBK02,KoHG08}.
In earlier work \cite{chord}, we used a model of churn similar to 
the one used here, to analytically 
estimate the performance of the Chord DHT under churn. 
The protocol studied here is far simpler and so we are able to model it
at a much finer level of granularity than \cite{chord} where we had to
treat lookups as atomic.

The effects of churn on network connectivity have been studied by
Leonard et al \cite{LeonardRL05}. They relate node lifetime distributions
to the likelihood that nodes get isolated, and estimate the
probability of p2p system partitioning. Kong et al \cite{KongBR06}
study reachability properties of DHT systems and obtain analytic
models of connectedness in terms of system size and node failure
probability. Wu et al \cite{WuTN07} model different DHT lookup mechanisms
and evaluate analytically their performance under churn. Several authors
\cite{qiu2004modeling,kumar2007stochastic} use stochastic fluid models
to evaluate the performance of p2p streaming
systems such as BitTorrent.

We know of few prior analytical performance analyses of tree-based protocols
under churn in the literature. In the context of multicasting,
van Mieghem \cite{vanMieghem05} investigates the stability of multicast
shortest-path trees by estimating the expected number of link changes in the
tree after a multicast group member leaves the group. The analysis is done
for random graphs and $k$-ary trees.
Lower bound results have been obtained for
Demmer and Herlihys Arrow protocol \cite{DemmerH98,KuhnW04}, and
the average case communication cost of Arrow is studied in
\cite{PelegR99}, but only for the static case. 

Other authors have used analytic methods to model the effects of
time-varying aggregation data, cf. \cite{prieto2007gap,jin2006lifetime,%
enachescu2005scale,oswald2008tight}, but again only for the case of
static networks.

\paragraph{Structure of Paper.}
Section \ref{sec:model} describes our network model and Section
\ref{sec:proto} an outline of the GAP aggregation protocol. In Sections
\ref{sec:nx} and \ref{sec:aggr} we construct a stochastic model for the
number $N_x$ of nodes with a certain distance $x$ to the root in a system
with a given rate of churn, 
and an estimated average partial aggregate $A_x$ held by a node at
distance $x$. Section \ref{sec:eval} presents the simulation results and we
end by discussing the implications of our results. Further details of the
calculations are relegated to the appendix.

\section{Network Model}
\label{sec:model}%
We model the network as an Erd\"{o}s-R\'enyi 
random graph $G(t)=(V(t),E(t))$ with Poisson-arrival node join and
(per node) failure rates $\lambda_j$ and $\lambda_f$ as described above. 
Nodes join with an a priori Poissonian degree distribution 
$Q(k)= e^{-\overline\lambda} \frac{{\overline\lambda}^k}{k!}$, 
where $\overline\lambda$ is the average degree of the new joinees,
% $k\in\Nat$ with associated random variable $Q(k)$, 
and attach themselves to nodes
that are already present in $G(t)$ with uniform probability. Node
failures/deletions remove a node $v$ chosen uniformly at random from $V(t)$ 
along with all edges attached to $v$.
We state the following observation without proof (cf. \cite{moore2006exact}).

\begin{proposition}
\label{100204.1}
Assume a network following the above conventions, and
let $\lambda_j=N \lambda_f$.
\begin{enumerate}
\item The a posteriori degree distribution $P_k(t)$ of $G(t)$ is
time independent and identical to $Q(k)$.
\item In expectation, the network size (number of nodes) of $G(t)$ is $N$.
\end{enumerate}
\end{proposition}

\section{The Protocol}
\label{sec:proto}
The protocol we analyze is based on the BFS algorithm of Dolev, Israeli, and
Moran \cite{dolev1993self}.
The graph has a distinguished root node $v_0$ which never fails, and which
executes a program slightly different from the others.

A node $v\in V(t)$ has three associated registers $\Level_v$, $\Aggregate_v$
and $\Parent_v$  holding the following values:
\begin{itemize}
\item $\Level_v\in\Nat$ holds $v$:s {\em level}, i.e. $v$:s 
current belief in the
number of hops needed to reach $v_0$.
\item $\Aggregate_v\in\Nat$ holds $v$:s {\em partial aggregate}: In this
case $v$:s current belief in the size of the subtree rooted in $v$.
\item $\Parent_v\in\bigcup\{v'\in V(t')\mid t' < t\}$ indicates the
node (which may no longer exist) which $v$ believes to be its parent in
the global aggregation tree.
\end{itemize}
A {\em configuration} is a graph $G(t)$ along with assignments to the three
registers for each node in $V(t)$.
We write $\Level_v(t)$ when we want to refer to the value of $\Level_v$
at time $t$ and similarly for the other registers.

Each node randomly and independently executes \textit{updates}, which are
Poisson distributed events with rates $\lambda_g$. Let $\Neighbours_v(t)$ be the neighbours of 
$v$ at time $t$. When one node $v$ executes an update at time $t$   
its registers are updated as follows:
\begin{itemize}
\item $\Level_v := \min\{\Level_{v'}\mid v'\in\Neighbours_v(t)\} + 1$,
where $\min(\emptyset) = \infty$ and 
$\infty+1=\infty$.
\item $\Aggregate_v := \Sigma\{\Aggregate_{v'}\mid v'\in\Neighbours_v(t) \wedge \Parent_{v'}=v \} + 1$.
\item $\Parent_v$ is some $v'\in\Neighbours_v(t)$ with minimum level.
If $\Neighbours_v(t) = \emptyset$ then $\Parent_v=v$.
\end{itemize}

Upon joining, a node $v$ updates its level and parent registers
as above, and initializes its aggregate register to $1$. We say
that a node $v$ {\em joins at level $x\in\omega$}, or, {\em joins at
defined level}, if, upon completed initialization,
$\Level_v = x \neq\infty$.

The root node updates only its aggregate register, as above, and initializes
$\Level_{v_0}$ to $0$ and $\Parent_{v_0}$ to $v_0$.

For a fixed expected graph size $N$, the initial configurations at 
time $t=0$ are obtained from 
a graph $(v_0,\emptyset)$ by nodes joining the graph according to the network
model and join protocol above, until the graph has reached size $N$.

It is not hard to see that in the static case, this algorithm converges in 
expected time $\lambda_g^{-1}$ times the diameter of the 
network. In fact, the
algorithm is self-stabilizing \cite{dolev1993self} in the sense that convergence
is independent of how the registers are initialized.

\section{The $N_x$ Model}
\label{sec:nx}
To estimate the long term average population
at level $x$ we use the following stochastic variables:
\begin{itemize}
\item $N_x(t)$ estimates the number of nodes with level assignment 
$x\in\omega$ at time $t$.
The level assignment may be correct in the sense that the node has a neighbour
with minimum level assignment $x-1$ (if it is not the root), or it may 
be incorrect,
in which case the level assignment will change upon an update.
\item $\Nxus(t)$ estimates the number of nodes at level $x$ with incorrect
level assignment. We call these nodes {\em unstable}. Clearly this is a subset of $N_x(t)$
\item $N^y_x(t)$ estimates the number of nodes with level $x$ and with minimum
neighbour level $y$. Upon performing an update, this node moves to level $y+1$.
\end{itemize}

\begin{corollary}
\[
\begin{array}{c}
N_0(t) = 1, \ \ N_{x+1}(t) = N^x_{x+1}(t) + \Nus_{x+1}(t),\ \ 
\Nus_0(t) = 0,\\[2pt]
\Nus_1(t)=0, \ \ \Nus_{x+1}(t) = \displaystyle\sum_{x'\neq x}N^{x'}_{x+1}(t)
\end{array}
\]
\end{corollary}
In this analysis, we are mainly interested in 
expectations (time averages) in the steady state. 
So  in most of what follows, we discard  the time
parameter and, for ease of notation, denote $E[N_x]$ simply by $N_x$. 

From the join protocol and in the large $N$ limit, the probability 
of a node joining level 
$x$ is denoted $p_{min}(x)(t)$:
\begin{equation}
p_{min}(x-1)(t) = \displaystyle\sum_k Q(k) [(1- 
\displaystyle\sum_{i=0}^{x-2}\frac{N_i(t)}{N})^k - 
(1- \displaystyle\sum_{i=0}^{x-1}\frac{N_i(t)}{N})^k]
\end{equation}
This is the probability that at least one of the nodes' $k$ links 
is connected to a node at level $x-1$ while 
no link has a connection to a lower level. This quantity is averaged over $k$.

The gain-loss terms for $N_x(t)$ are determined as follows:
\begin{equation}
\label{100203.6100203.6}
N_x(t + \Delta t) \rightarrow \left\{
\begin{array}{rl}
N_x(t)+1 & \text{w.p. } \lambda_{j} p_{min}(x-1)(t) \\
N_x(t)-1 & \text{w.p. } \lambda_{f} N_x(t) \\
N_x(t)+1 & \text{w.p. } \lambda_{g} \sum N_{x'}^{x-1}(t)\,\, x'\neq x \\
N_x(t)-1 & \text{w.p. } \lambda_{g} N_x^{us}(t) \\
\end{array} \right.
\end{equation}
The first (gain) term quantifies the change when a node joins and the 
second (loss) term accounts for node failures. 
The third term is the probability that a node with an incorrect level 
assignment actually has its lowest connection 
at level $x-1$ and so is liable to join level $x$ upon an update. 
This constitutes an influx into level $x$ due to node updates.
The last term 
quantifies the probability that a node with incorrect level $x$  
updates its level and is removed from level $x$.
This term thus estimates the outflux from level $x$ due to node updates.

The above gain-loss terms translate to a time-dependent equation for $N_x$, but in all that follows we are
only interested in the steady state. That such a steady state exists (and is unique) for this process
is guaranteed by the properties of continuous time Markov processes \cite{mitzenmacher2005probability}.

In steady state, the above gain-loss terms make the following prediction for $N_x$

\begin{equation}
\label{100203.6100203.7}
\frac{N_x}{N} = p_{min}(x-1) + r\left(\Sigma_{x' \neq x} \frac{N_{x'}^{x-1}}{N} - \frac{\Nus_x}{N} \right)
\end{equation}
The term inside the brackets is 
the difference of the {\it influx} of nodes after an update event into level 
$x$,  from an {\it outflux} of unstable nodes out of $x$.
On general grounds, this term may be expanded in a series 
$\sum y_k/r^{k}$. The first term in the sum which is of order $1/r$, 
when multiplied by $r$ will 
be independent of $r$ and hence to leading order, 
we expect $N_x$ to be independent of $r$. However if the term inside
the brackets is non-zero, we would have sub-leading order corrections, 
and hence an $r$-dependence especially for low $r$.
We make the important simplifying assumption that this does not happen:
\begin{assumption}\label{influxoutflux}
\label{100203.1}
In expectation,  the influx of nodes into level $x$, 
$\sum_{x' \neq x} N_{x'}^{x-1}$, 
is equal to the outflux of nodes from level $x$, $\Nus_x$.
\end{assumption}
Below we discuss why and under what conditions 
assumption \ref{influxoutflux} might be valid, but, if the 
assumption {\em is} valid
then, from Eq.  \ref{100203.6100203.7}, we have
$N_x/N=p_{min}(x-1)$. Thus if 
influx and outflux are in equilibrium, the level distribution is independent 
of $r$ and equal to the a priori distribution. 
Indeed this is confirmed by the simulation
results, Figure \ref{fig:nx}, with caveats related to the possibility
of nodes disconnecting from the root.

The main reason why assumption \ref{100203.1} might fail to hold is
disconnection. The probability of disconnection decays exponentially in the
degree $\overline\lambda$. For this reason we conjecture that assumption \ref{100203.1} holds
with probability overwhelming in $\overline\lambda$ and $N$. 
 
Disconnection may happen for several reasons. 
Most interesting is the count-to-infinity
problem where, upon disconnecting, a cluster of nodes' level registers
begin to diverge. In this case we do not expect
influx and outflux to be
in equilibrium, as the lowest levels in that case have only outflux,
and the highest only influx. In the absence of disconnection,
however, we expect
join and failures at any level $x' < x $  to have opposing, but equal
effects on the (eventual) instability at level $x$.
If this were not the case, there would be a continuous accretion 
or depletion of nodes from level $x$, something which would contradict
both intuition and experimental evidence.

In what follows, we make use of assumption \ref{influxoutflux}
to estimate both the size of the perceived subnetwork rooted at 
a node at level $x$, as well as the number of unstable nodes at level $x$.

\section{Aggregation}
\label{sec:aggr}
We now turn to the stochastic variables $M_x(t)$, which is  the aggregate 
held by a node at level $x$. $A_x(t) = M_x(t) N_x(t)$  estimates the
total aggregate held at level $x$. We can write the following gain-loss terms for this quantity

\[
A_x(t + \Delta t) \rightarrow \left\{
\begin{array}{ll}
A_x(t)- M_x(t) & \text{w.p. } \lambda_{f} N_x \\
A_x(t)+ 1 & \text{w.p. } \lambda_{j}p_{min}(x-1) \\
A_x(t)- M_x(t) & \text{w.p. } \lambda_g N_x^{us} \\
A_x(t) + M_{x'}(t) & \text{w.p. } \lambda_g \sum N_{x'}^{x-1} \,\, x'\neq x\\
A_x(t)- M_x(t) + 1 + \frac{\Ns_{x+1}}{N_x} M_{x+1}(t) & 
\text{w.p. } \lambda_g N_x^{s} \\
\end{array} \right.
\]
We account for all the processes, which in an infinitesimal unit of time, could change $A_x$: Level $x$ loses
$M_x$ in expectation upon failure of a node at level $x$; upon a join,
$A_x$ is increased by one; upon an update of an unstable node leaving level
$x$, $M_x$ is lost (in expectation); upon an update of an unstable node
at level $x'$ entering $x$, $M_{x'}$ is gained; finally, upon an update
of a stable node, the current expected aggregate is lost, and the sum
of the aggregates of the children plus one is gained. For the last term we
estimate the expected number of children of a node at level $x$ by the quantity
$d_x = \Ns_{x+1}/N_x$ where $\Ns_x = N_x - \Nus_x$ represents the number of
stable nodes at level $x$. This is a good estimate but not exact, as
there might be nodes at level $x+1$ which have a better choice of parent 
but do not as yet know this (not having executed an update). In this
case, from the protocol, they would still send their aggregate to 
their outdated parent. However we ignore this effect in the theory.

Using $a_x = A_x/N = M_x N_x/N$ the steady state equation
is :

\begin{equation}
a_x(1+r) = p_{min}(x-1) + r \displaystyle\sum_{x' \neq x} \frac{M_{x'} 
N_{x'}^{x-1}}{N} + r \frac{N_x^s}{N} + r a_{x+1} 
\frac{N_x^s}{N_x}\frac{N_{x+1}^s}{N_{x+1}}
\end{equation}
Further simplification and assumption 
\ref{influxoutflux} produces the following
recursion relation:
\begin{equation}
\label{100203.5}
a_x = p_{min}(x-1) + \frac{r}{r+1} (a_{x+1} \frac{N_x^s}{N_x}\frac{N_{x+1}^s}{N_{x+1}} + \displaystyle\sum_{x' \neq x} (M_{x'} - 1) \frac{N_{x'}^{x-1}}{N} )
\end{equation}
To solve (\ref{100203.5}), we need to estimate 
$\Ns_x$ (or equivalently, $\Nus_x$), which is a function of $r$. 
We present this calculation in the appendix. The calculation 
involves estimating the outflux
out of level $x$ due to all the processes which might make a node at level
$x$ unstable: namely, the node has a single connection to the earlier level 
which fails; or the single connection is itself unstable and on an  
update makes the level-$x$ node unstable; or a new joinee 
or one of the neighbors of the node provides a better
choice of parent. We show from comparison to simulations, that 
our estimate is quite good for
all the values of the parameters considered.

To solve Eq. (\ref{100203.5}), we also need to estimate the last term
which involves terms $N_{x'}^{x-1}$. 
To this end we make
the simplifying assumption (consistent with assumption \ref{influxoutflux}), 
that for $x'\neq x$, $N_{x'}^{x-1}$, the influx to $x$ from $x'$, 
is roughly equal to 
the population at level $x'$ times the ratio of outflux of level $x$ to the
entire population, i.e. 
\begin{equation}
\label{100203.3}
x'\neq x \Rightarrow N_{x'}^{x-1} \approx \frac{N_x^{us}}{N} N_{x'}\ .
\footnote{In fact the use of $N$ here is a slight overestimation, 
as levels $0$,$1$, and $x$ itself need to be excluded.}
\end{equation}
Note that this assumption can itself be used to estimate $\Nus_x$ and
gives a far worse estimate than the one we derive in the appendix. 
However this assumption helps  in making the last term in Eq.
(\ref{100203.5}) tractable.

Using (\ref{100203.3})
Eq. (\ref{100203.5}) can then be simplified to
\begin{equation}
\label{102003.4}
a_x = p_{min}(x-1) + \frac{r}{r+1} (a_{x+1} 
\frac{N_x^s}{N_x}\frac{N_{x+1}^s}{N_{x+1}} + 
\frac{N_x^{us}}{N}
\displaystyle\sum_{x'\neq 0,1,x}(a_{x'}-p_{min}(x'-1))
\end{equation}
This equation can be solved numerically, once we have an expression for
$\Nus_x$. Since the L.H.S has a sum over $a_{x'}$, which is the quantity we want to compute,
we solve the equation iteratively, to get a self-consistent solution.

\section{Validation Through Simulation}
\label{sec:eval}
We evaluate the model developed in Sections \ref{sec:nx} and \ref{sec:aggr} 
through simulation studies using a discrete-event simulator. We simulate the 
protocol described in Section \ref{sec:proto} executing on a network modeled 
as described in Section \ref{sec:model}. The events we simulate are nodes 
joining, nodes failing and the execution of protocol cycles on a node. (The 
protocol does not distinguish between a node failing and a node leaving the 
network.)

When a join event occurs, a new node is created on the network graph, links 
from this node to other nodes of the network graph are created, the node's 
registers are initialized following the join protocol described in Section 
\ref{sec:proto}, and the node executes a protocol cycle. When a fail event 
occurs, the node is removed from the network graph, together with the links 
that connect it to other nodes in the graph. Finally, during the execution 
of a protocol cycle on a node, the registers of the node are updated as 
described in Section \ref{sec:proto}, whereby the node changes its parent only if a 
neighbor with a lower level than its current parent exists. 

During a simulation run, we periodically sample the metrics $N$, $N_x$, $A_x$ 
and $N^{us}_x$. Each measurement point on the graphs in this section 
corresponds to values averaged over at least a 1,000 such samples. 

All runs simulate the system in steady-state. The network 
graph for the simulation is a random graph with $N$ nodes with
a probability $p$ that any pair of these nodes is connected. 
($p=\overline\lambda/N$ where $\overline\lambda$ is the average node degree.)

We perform simulation runs for all combinations of the following parameters: 
number of nodes ($N$): 1,000, 10,000, 100,000; average node degree ($pN$): 4, 
6, 8; $r$: 1, 10, 100, 1,000. Runs for a single simulation 
can be very long for large networks and small $r$. For instance, for 
$N=100,000$ and $r=1$, the simulation takes about a month on a single server 
to collect 1,000 samples. 

Using these simulation results, we validate the following predictions of our 
models, for steady state conditions: the level distribution, the accuracy of 
the counting protocol and the distribution of unstable nodes. The first 
two predictions are discussed in this section, while the last is discussed in the appendix.

\paragraph{Validation of Level Distribution ($N_x/N$).}
In this series of experiments, we validate the $N_x$ model (see Section 
\ref{sec:nx}) under steady-state conditions. 

Specifically, we 
compare the simulation results with the prediction of equation 
(\ref{100203.6100203.7}).

\begin{figure}[!ht]
\begin{center}
  %\centering
\subfigure[]{
  \resizebox{7cm}{!}{\includegraphics[]{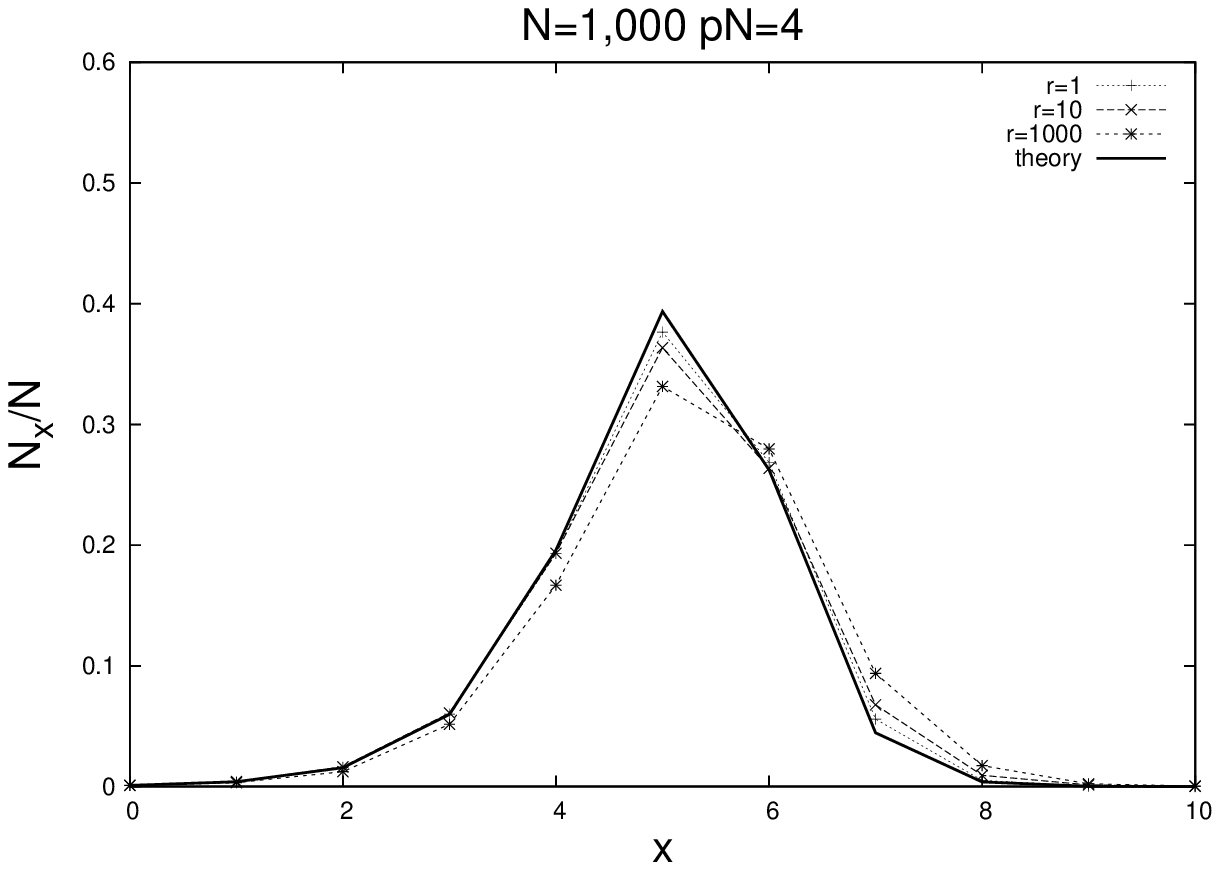}}
\label{fig:nxN1kd4}
}
\subfigure[]{
  \resizebox{7cm}{!}{\includegraphics[]{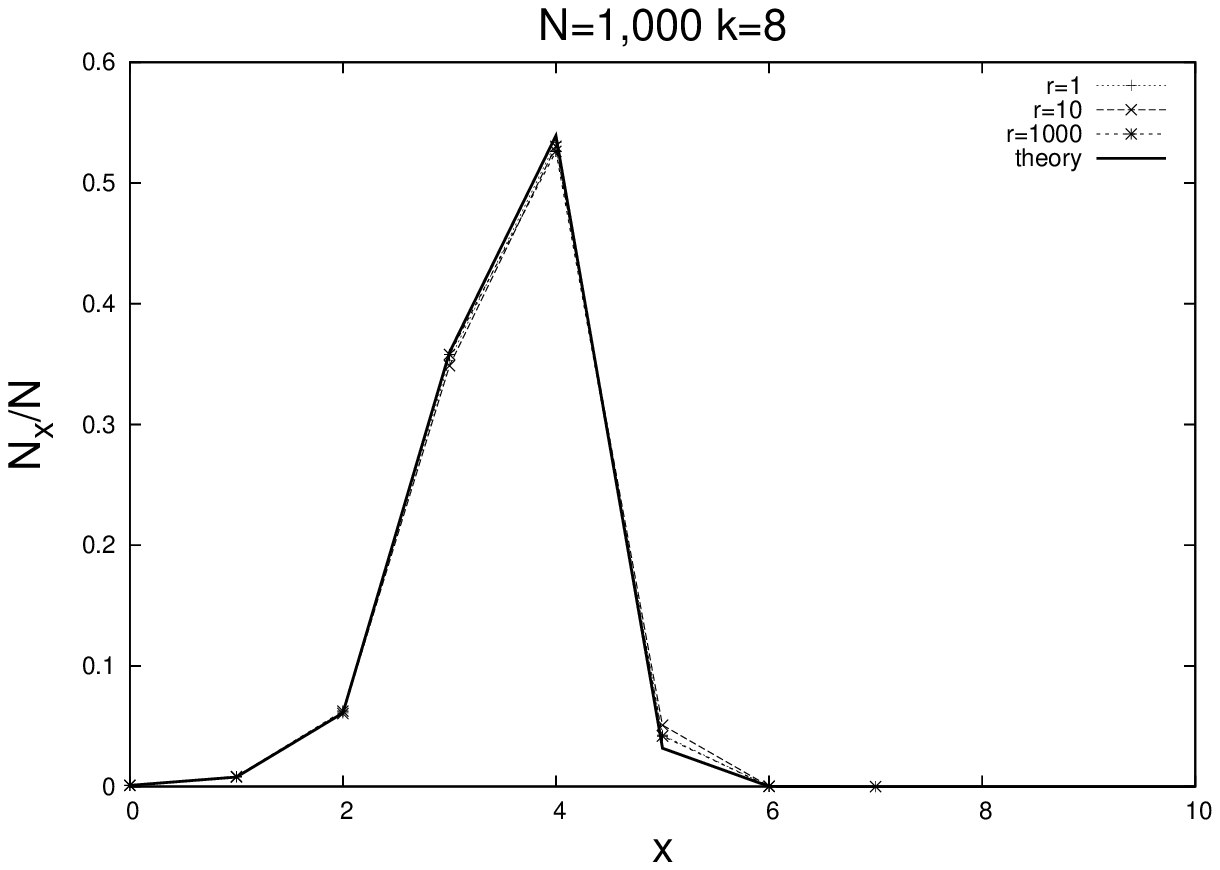}}
\label{fig:nxN1kd8}
} \\
\subfigure[]{
  \resizebox{7cm}{!}{\includegraphics[]{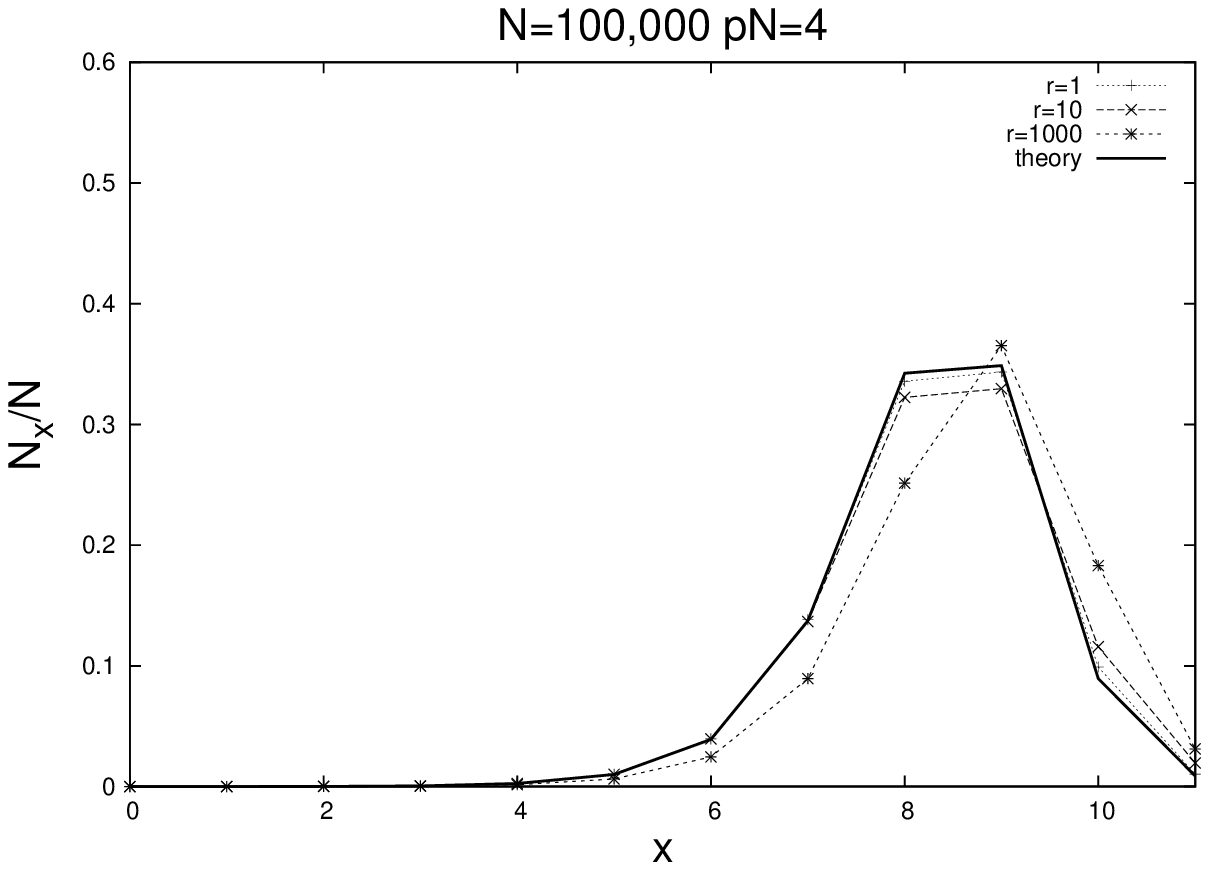}}
\label{fig:nxN100kd4}
}
\subfigure[]{
  \resizebox{7cm}{!}{\includegraphics[]{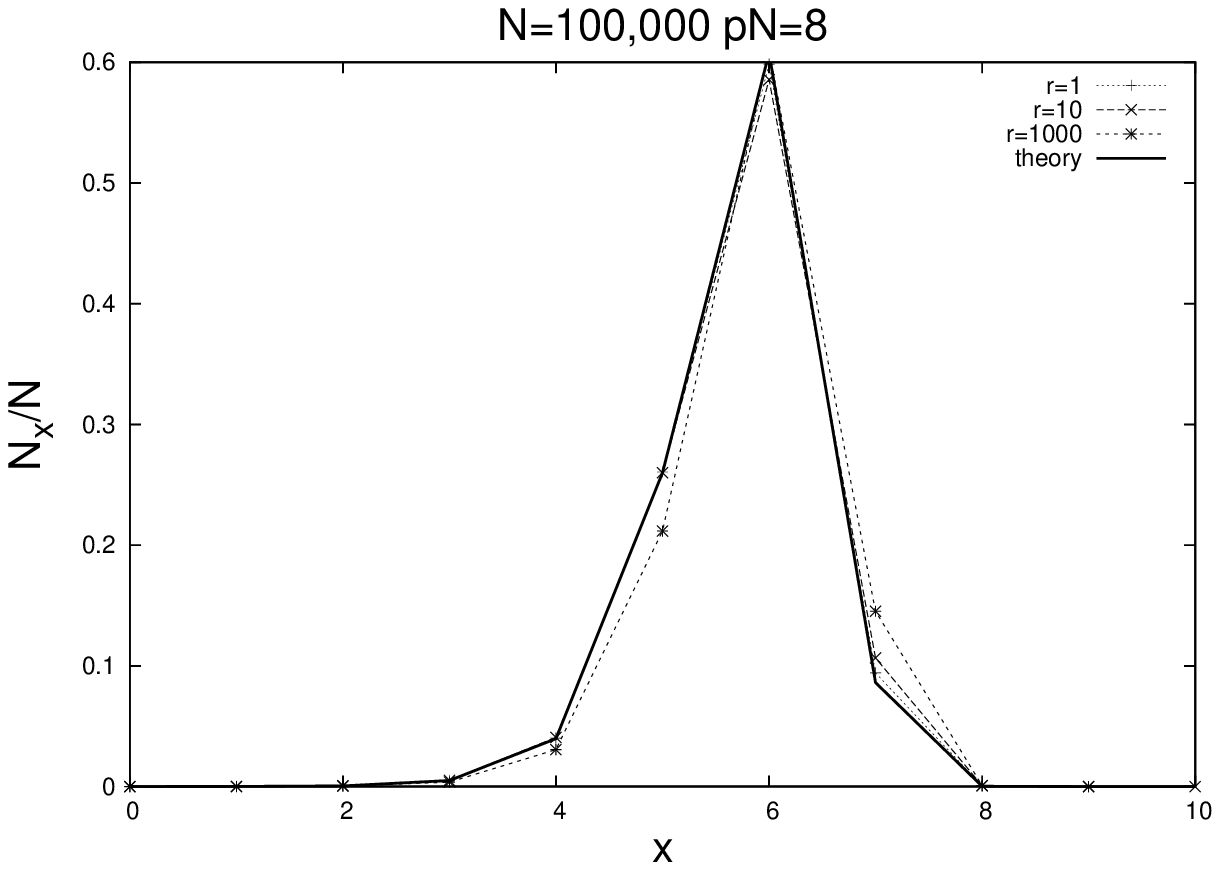}}
\label{fig:nxN100kd8}
}
\end{center}
\caption{Comparing theory with simulation results: the level distribution $N_x/N$ as a function of network size, node degree $pN$ and $r$.}\label{fig:nx}
\end{figure}

Figure \ref{fig:nx} shows in a series of graphs, the numerical evaluation of 
equation (\ref{100203.6100203.7}), together with simulation results, for 
network sizes of 1,000 and 100,000, degrees of 4 and 8, and rates of 1, 10, 
1,000. 

We make the following observations. First, the model predictions fit the 
actual simulation results very well. As predicted, the distribution seems to 
be independent of the parameter $r$. Second, as expected, an increase of the 
network size, or a decrease of the node degree enlarges the average tree 
height.

\graphicspath{{./Final/}}
% NX

\paragraph{Validation of $A_x/N$.}
In this series of experiments, we validate the $A_x$ model (see Section \ref{sec:aggr}) under steady-state conditions. 

Specifically, we compare the simulation results with the prediction of equation (\ref{102003.4}).

\begin{figure}[!ht]
\begin{center}
\subfigure[]{
  \resizebox{7cm}{!}{\includegraphics[]{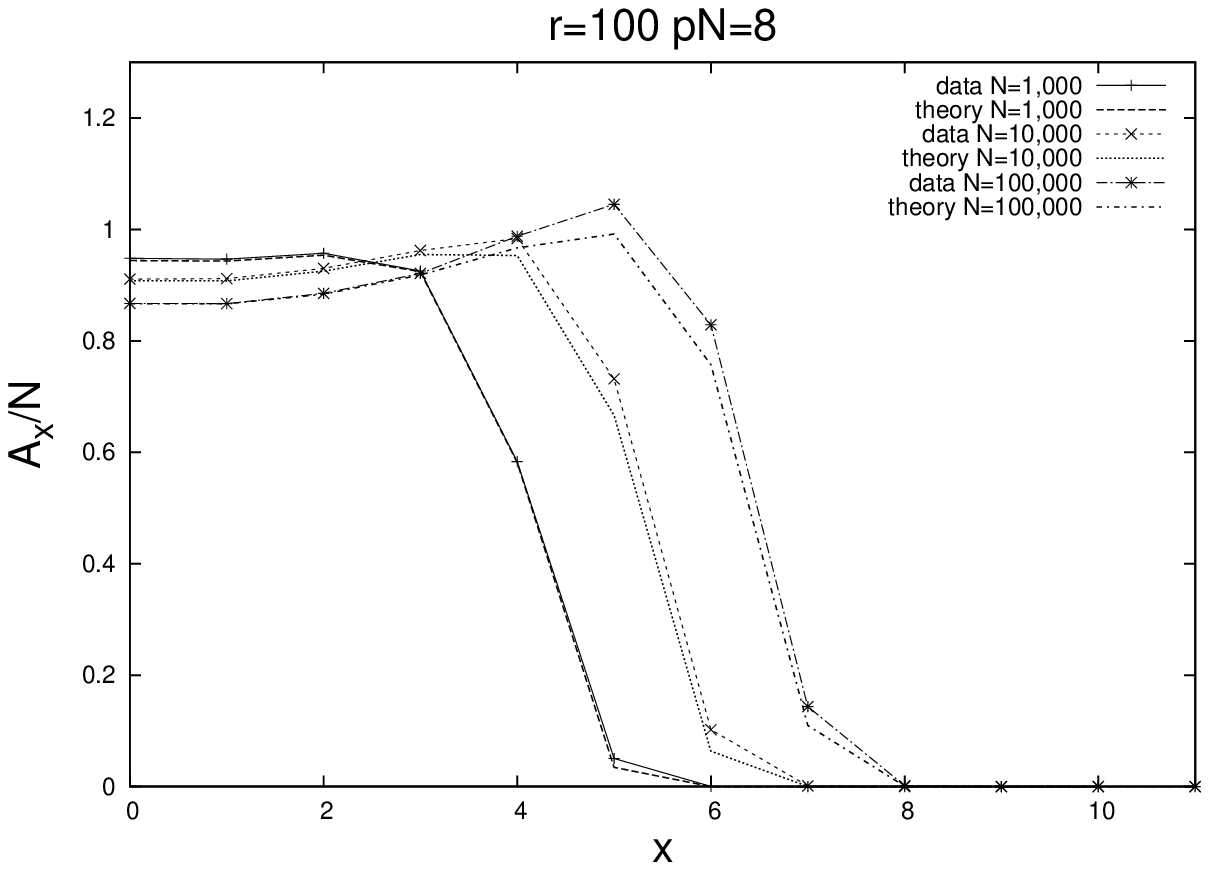}}
\label{fig:axN100d8}
}
\subfigure[]{
  \resizebox{7cm}{!}{\includegraphics[]{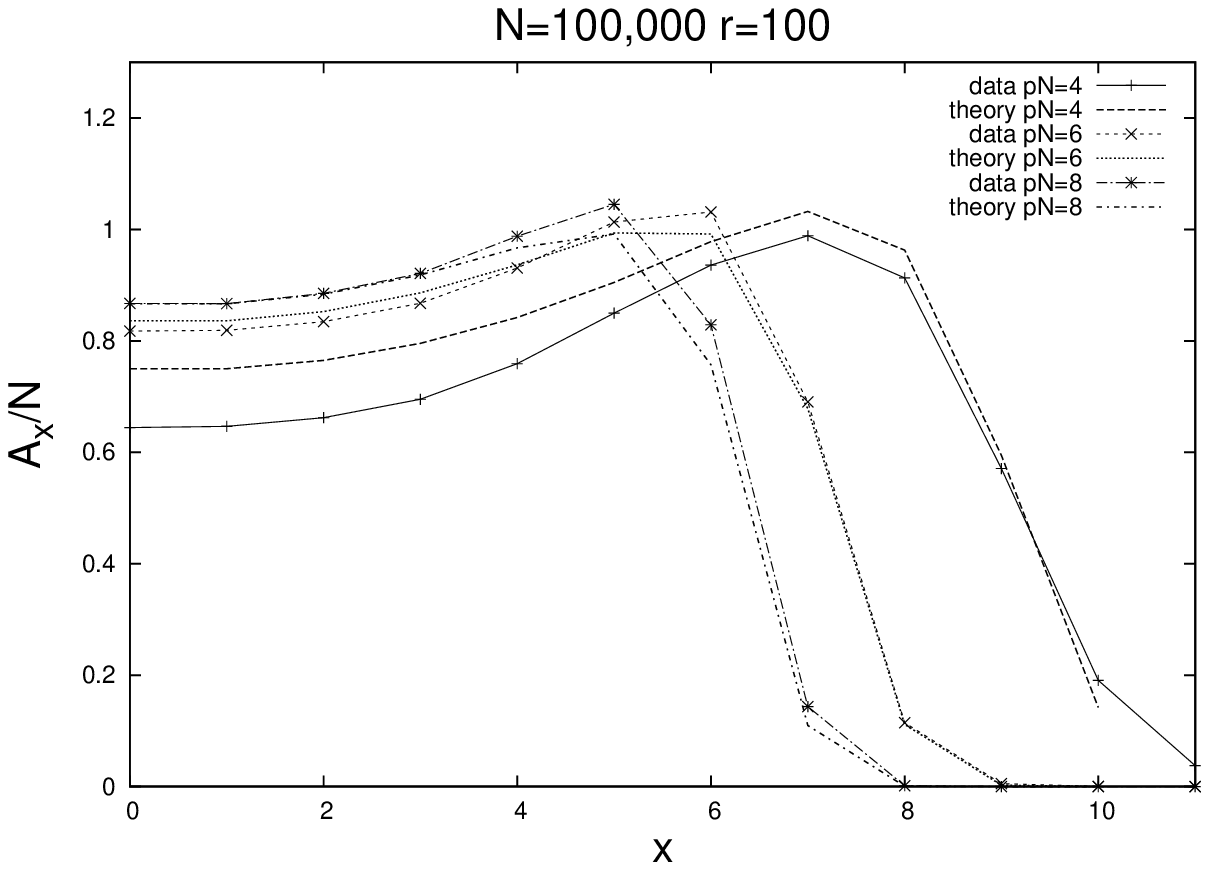}}
\label{fig:axN10d8}
} \\
\subfigure[]{
  \resizebox{7cm}{!}{\includegraphics[]{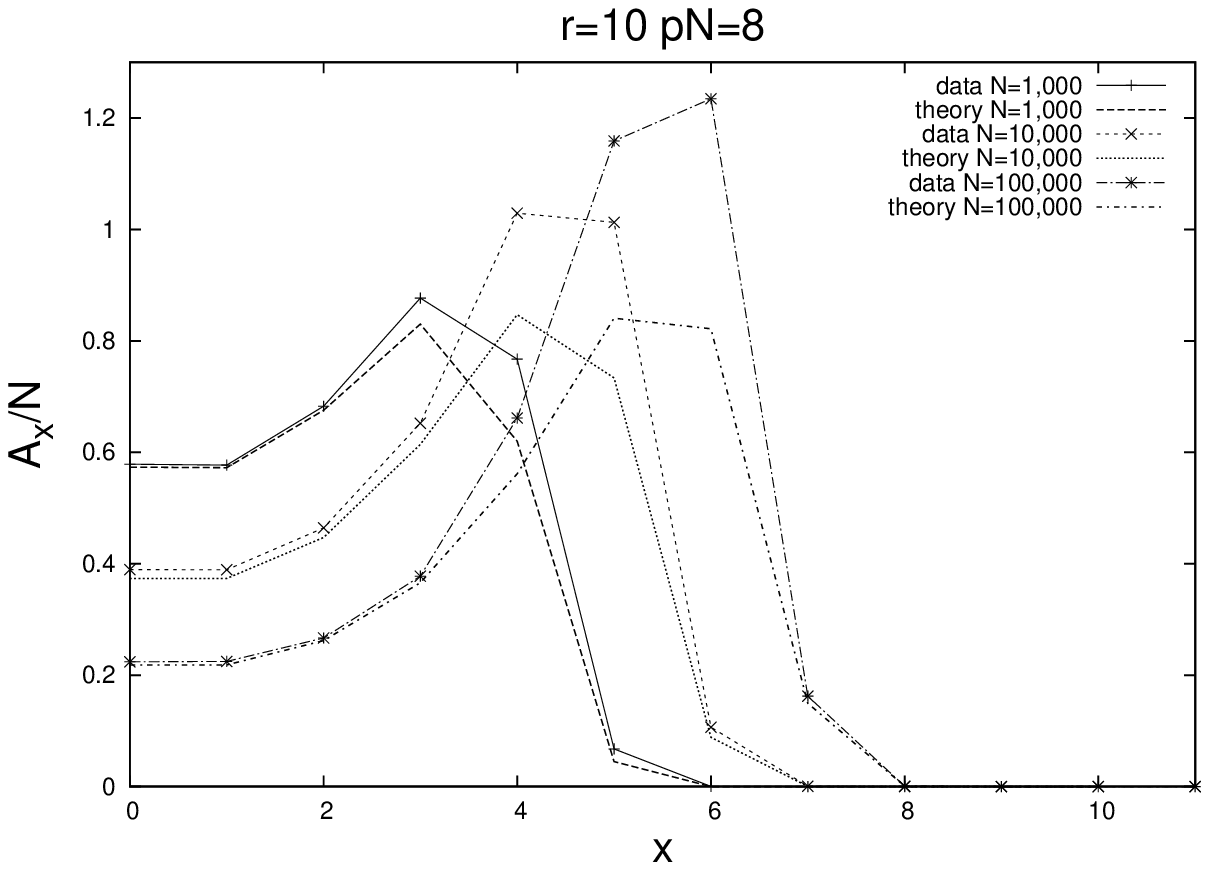}}
\label{fig:axN100000r100d8}
}
\subfigure[]{
  \resizebox{7cm}{!}{\includegraphics[]{Ax_r100_k8.eps}}
\label{fig:axN10000d8}
}
\end{center}
\caption{Comparing theory with simulation results: $A_x/N$ as a function of network size, node degree and $r$. For $x=0$, the curves show the accuracy of of the protocol for counting.}\label{fig:ax}
\end{figure}

Figure \ref{fig:ax} shows, in a series of graphs, the numerical evaluation of equation (\ref{102003.4}), together with selected simulation results, for network sizes of 1,000, 10,000, 100,000, degrees of 4, 6, 8, and rates of 1, 10, 100, 1,000.

These results suggest the following. First, the accuracy of the prediction
of the model increases with $r$ and $N$, and decreases with $x$. Second, we
observe that the accuracy of the counting protocol decreases with increasing
network size and, as expected, the accuracy increases with the degree of the
network graph and the parameter $r$.

Of specific interest to us is the accuracy of the counting protocol, which
depends on the prediction accuracy of $A_0/N$.  Table \ref{tbl:a0} gives the
relative errors of the model prediction of $A_0/N$ against the simulation
measurements for various parameters.  We notice that the predictions of the
model are good for large values of $r$. For instance,
for $r=100$, i.e., for systems with an average lifetime of about 100
protocol cycles per node,  the prediction error is at most 2\%, for 
network graphs with average node degree of 6 or 8. We can also see that for a
highly dynamic system with $r=1$, the error is much larger. To illustrate
the system dynamics, in the scenario with $r=1$, $pN=8$, and $N=100,000$,
the model predicts  $A_0=201$, while the simulation gives $A_0=122$, which
is hopelessly wrong for a counting protocol.

\begin{table}%
\begin{center}
\begin{tabular}{|c||c|c|c||c|c|c||c|c|c||c|c|c||}
\hline
& \multicolumn{3}{|c||}{$r=1$}
& \multicolumn{3}{|c||}{$r=10$}
& \multicolumn{3}{|c||}{$r=100$}
& \multicolumn{3}{|c||}{$r=1,000$}\\\hline
$pN$& $4$ & $6$& $8$
& $4$ & $6$& $8$
& $4$ & $6$& $8$
& $4$ & $6$& $8$ \\\hline
$N=10^3$ &
.03 & .05 & .07 & .02 & .00 & .01 & .02 & .00 & .01 & .03 & .00 & .00
\\\hline
$N=10^4$ &
.12 & .27 & .28 & .11 & .00 & .04 & .03 & .01 & .00 & .09 & .00 & .00
\\\hline

$N=10^5$ &
.37 & .56 & .65 & .58 & .10 & .00 & .16 & .02 & .00 & .04 & .00 & .00
\\\hline

\end{tabular}
\end{center}
\caption{Relative errors of model prediction of $A_0/N$ versus simulation
results in steady state, for network size $N$, average node degree $pN$ and
$r=\lambda_g/\lambda_f$.}
\label{tbl:a0}
\end{table}

\graphicspath{{./Final/}}
% AX

\section{Discussion}

\label{sec:conc}
The contribution of this paper is an important first step towards a general
theory of tree-based protocols under churn. We have shown that key
performance metrics of a simple, tree-based aggregation protocol can be
calculated systematically, and that these calculations are mostly in good
agreement with simulation results. In addition, the equations we have
derived can be solved numerically for any set of parameter values, to
predict the expected accuracy of the estimate at the root, which requires
only a few seconds on a regular laptop. (This is in contrast to a
simulation, which, depending on the parameter values of interest, can take
months on a single processor.) We believe that using such analytical tools
to predict performance metrics can be of invaluable assistance in  
understanding the
behavior of large-scale distributed information systems, and hence open up
exciting avenues for engineering new protocols that execute effectively in
dynamic environments.

As the reader surely has noticed, the analysis presented in this paper, is
not rigorous. Simplifying assumptions are made and theorems
or exact bounds are not proved. However, this method follows a
well-established practice in physics, namely that of deriving equations for
expected values and, in the case they are not solvable, simplifying these
while ensuring that global invariants (such as proposition ~\ref{100204.1} or 
assumption \ref{influxoutflux} ) are not violated. The simplifications of such equations 
usually  fall into three categories. The first involves simplifications that affect
the very form of the equation, by disregarding fluctuations that could lead to
terms with higher moments. Equations for these higher moments would then
possibly involve still higher moments, leading in most cases to an unsolvable
hierarchy of equations. Such an approximation is made, for instance, in Equation
~\ref{100203.5}, in assuming that $E[d_xM_{x+1}] = E[d_x]E[M_{x+1}]$. 
The second simplification appears in assumptions like those made in the estimation 
of $d_x$ or $N_{x'}^{x-1}$. This type of simplification is a less serious modification of
the theory than the first one, since a better estimate of these terms is all
that is required for more accurate predictions. The third simplification
involves using expressions, such as Equation. 1, which strictly hold only in the
large system limit. As we see from Figure \ref{fig:ax} our fits do get better as $N$
increases. Proving this rigorously, however, is a very hard task and not
addressed here.

\bibliographystyle{plain}
\bibliography{bib}

\section{Appendix}

\subsection{Estimation of $N_x^{us}$}

We present below the details of the calculation for $\Nus_x$. This calculation
goes through the usual steps of detailing gain/loss terms, finding appropriate 
expressions for these and solving the equation for the expectation value in the steady state.
As in the earlier estimates, we use both proposition ~\ref{100204.1} and 
assumption ~\ref{influxoutflux}. Specifically, we use assumption ~\ref{influxoutflux}
to justify our use of the a-priori distribution in determining 
quantities such as the fraction of nodes at level $x$, having only 
one connection to the level above. Note that we cannot (and do not) use the
a-priori distribution to determine the average number of connections that a 
node has to the level {\it below}. This is  a quantity which is set by the parameter $r$. But
we are consistent with both proposition ~\ref{100204.1} and 
assumption ~\ref{influxoutflux} in assuming that the average number of connections
to the level above (as well as the average number of connections within the same level) are
independent of $r$, and given by the a-priori distribution.

\paragraph{The $x \geq 2$ case.}
Since the root never fails, nodes at level 1 can never become
unstable. We thus need to estimate unstable nodes only for level $x \geq 2$
We obtain the following gain-loss terms for $\Nus_x$:

\begin{equation*}
N_x^{us}(t + \Delta t) \rightarrow \left\{
\begin{array}{rl}
N_x^{us}(t)-1 & \text{w.p. } \lambda_{f} N_x^{us}\\
N_x^{us}(t)-1 & \text{w.p. } \lambda_{g}N_x^{us}\\
N_x^{us}(t)+1 & \text{w.p. } \text{(A)}\\
N_x^{us}(t)+1 & \text{w.p. } \text{(B)}\\
N_x^{us}(t)+1 & \text{w.p. } \text{(C)}\\
\end{array} \right.
\end{equation*}
The probabilities (A)--(C) are determined as follows:
\begin{itemize}
\item[(A)] This is the probability that a single connection at level $x-1$ fails or becomes
unstable. If the connection fails, by definition the node at level $x$
automatically becomes unstable. If the connection becomes unstable, the level-$x$ node
becomes unstable {\it only} after the parent has perfomed an update, which explains why the second term 
in the following equation is multiplied by $r$.  Introducing the notation
$\Ns_x(1)$ and $N_x(1)$ for the number of stable nodes at level $x$ with exactly one parent
and the number of nodes at level $x$ with exactly one parent, respectively, 
(A) is then
$\Ns_x(1)(1 + r \Nus_{x-1}/N_{x-1})/N$. 
We can write this as:
\begin{equation}
[\frac{N_x^s(1)}{N_x^s}][\frac{N_x^s}{N_x}][\frac{N_x}{N}][1+r\frac{N_{x-1}^{us}}{N_{x-1}}]
\end{equation}
Using the a priori distribution as mentioned earlier, we take
\begin{equation}
\frac{\Ns_x(1)}{\Ns_x} \approx \frac{N_x(1)}{N_x} 
\equiv \alpha_x = \overline \lambda \frac{N_{x-1}}{N_x} 
e^{-\overline \lambda \sum_{i=0}^{x-1} \frac{N_i}{N}}\ .
\end{equation}
\item[(B)] This is the probability that a new joinee with better level connects 
to a stable node at level $x$. If such an event occurs, then the node at $x$ has a better choice of parent then
the current one at $x-1$, and hence becomes unstable.
Using the a-priori distribution, We take (B) as
\begin{eqnarray*}
\lefteqn{
\frac{N_x^s}{N} \left[\frac{{\overline \lambda}^2}{N} e^{-\overline 
\lambda / N} + \overline \lambda \displaystyle\sum_{m=1}^{x-3}
[e^{-\overline \lambda \sum_{i=0}^{m-1}\frac{N_i}{N} }  
- e^{-\overline \lambda \sum_{i=0}^{m}\frac{N_i}{N} }] \right] } \\
 & = &
\frac{N_x^s}{N}\left[\frac{{\overline \lambda}^2}{N} e^{-\overline 
\lambda / N} + \overline \lambda [e^{-\overline \lambda / N }  - 
e^{-\overline \lambda \sum_{i=0}^{x-3}\frac{N_i}{N} }]\right]
\end{eqnarray*}
This is the average number of connections to stable nodes at level 
$x$ that a new joinee with a parent at level 0,1 ... up to $x-3$, has. 
Such a new joinee will then make the node at level $x$ change its level,
when it updates. 
\item[(C)] The probability that a
neighbor of a node at level $x$ provides a better choice of parent, i.e. that at
least one of the neighbor nodes is unstable with a better parent at level
strictly less than $x-2$. When such a neighboring node performs an update,this makes the node
at level $x$ unstable.

We first estimate $\lambda_x$, 
the average degree of a node at level $x$ using the a 
priori distribution. In order 
to not overcount the
case of a single parent (case (A)), we consider {\it all} neighbours of a node
with more than one connection to the parent level and all neighbours {\it excluding} the parent, 
for a node with only one connection
to the parent level . $\lambda_x$ then takes the value
\begin{equation}
\frac{
  \overline \lambda
  (
  1 - \displaystyle\sum_{i=0}^{x-2}\frac{N_i}{N}
  )
  (
  e^{-\overline \lambda \sum_{i=0}^{x-2}\frac{N_i}{N}}
  ) - \overline \lambda
  (
  1 - \displaystyle\sum_{i=0}^{x-1} \frac{N_i}{N}
  )
  (
  e^{-\overline \lambda \sum_{i=0}^{x-1} \frac{N_i}{N}}
  ) - \overline \lambda \frac{N_{i-1}}{N} e^{-\overline \lambda \sum_{i=0}^{x-1} N_i/N}
  }{p_{min}(x-1)
}
\end{equation}

Secondly, we estimate the probability of a neigbouring node
being unstable with a parent at level $1, \cdots,x-3$. We bound this
probability by the sum $\sum_{i=2,x-2} \frac{N_i^{us}}{N}$. 
This approximation is the same as is made in Eq. \ref{100203.3}.
As a consequence we get 
(C)$\mbox{} = r \lambda_x \sum_{i=2}^{x-2}\frac{N_i^{us}}{N}$
\end{itemize}
Putting the above together we obtain:
\begin{equation}
\label{eq:nxus}
\frac{N_x^{us}}{N_x} = \frac{D + E + F}{1 + D + E}
\end{equation}
where
\begin{eqnarray*}
D & = & \frac{\alpha_x}{1+r}(1 + r\frac{N_{x-1}^{us}}{N_{x-1}}) \\
E & = & [{\overline \lambda}^2 e^{-\overline \lambda /N} + \overline \lambda(e^{\overline \lambda /N} - e^{-\overline\lambda \sum_{i=0}^{x-3}\frac{N_i}{N}}]\frac{1}{1+r} \\
F & = & \frac{r}{1+r}\lambda_x \displaystyle\sum_{i=2}^{x-2}\frac{N_i^{us}}{N}
\end{eqnarray*}

\begin{figure}[!ht]
\begin{center}
\subfigure[]{
  \resizebox{7cm}{!}{\includegraphics[]{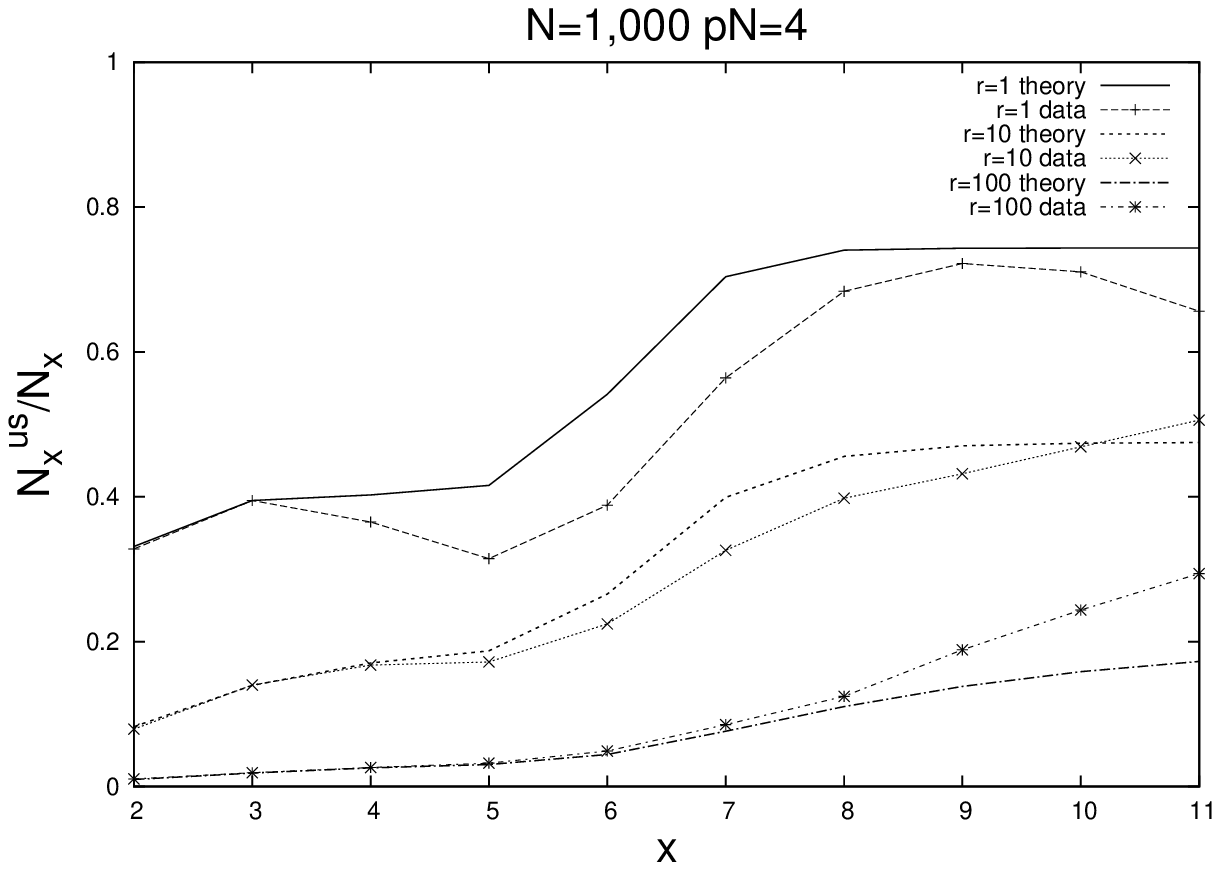}}
\label{fig:nxusN1kd4}
}
\subfigure[]{
  \resizebox{7cm}{!}{\includegraphics[]{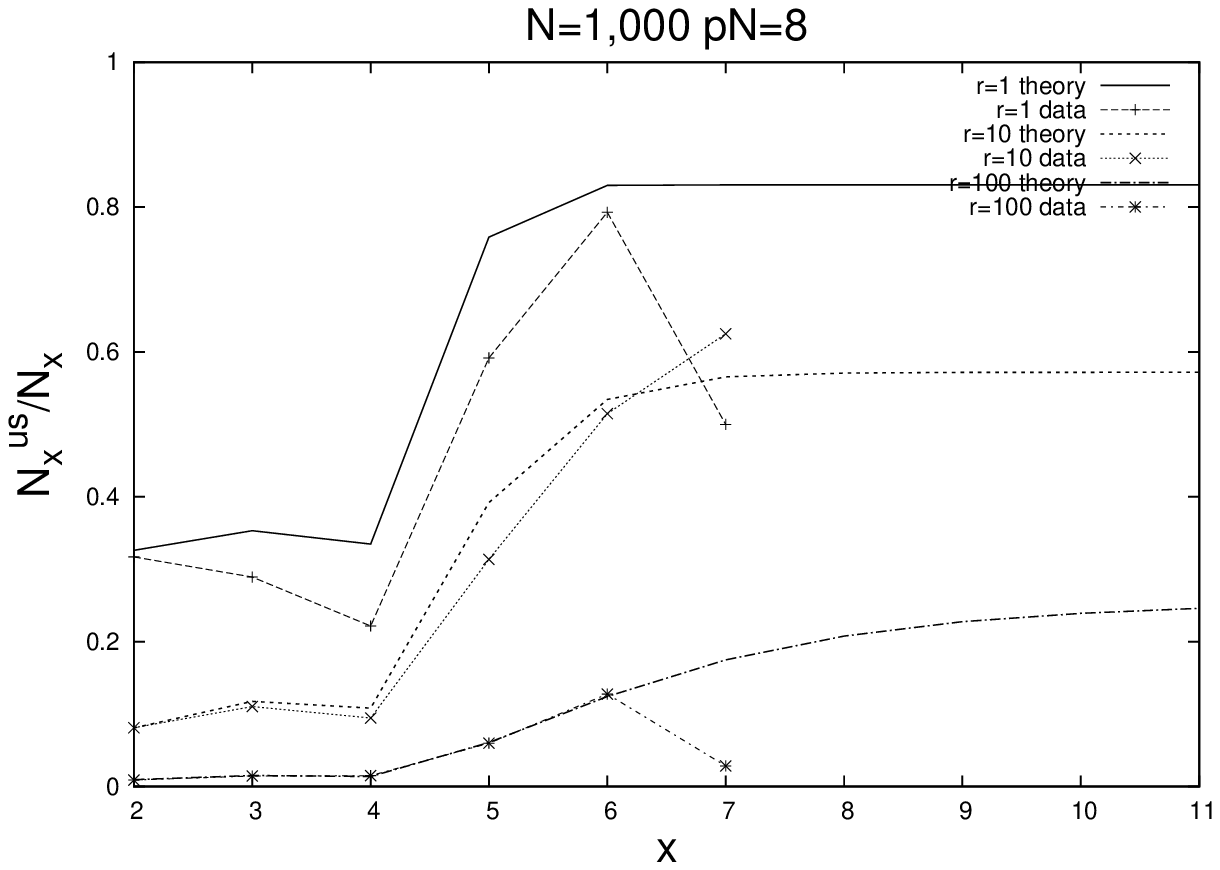}}
\label{fig:nxusN1kd8}
} \\
\subfigure[]{
  \resizebox{7cm}{!}{\includegraphics[]{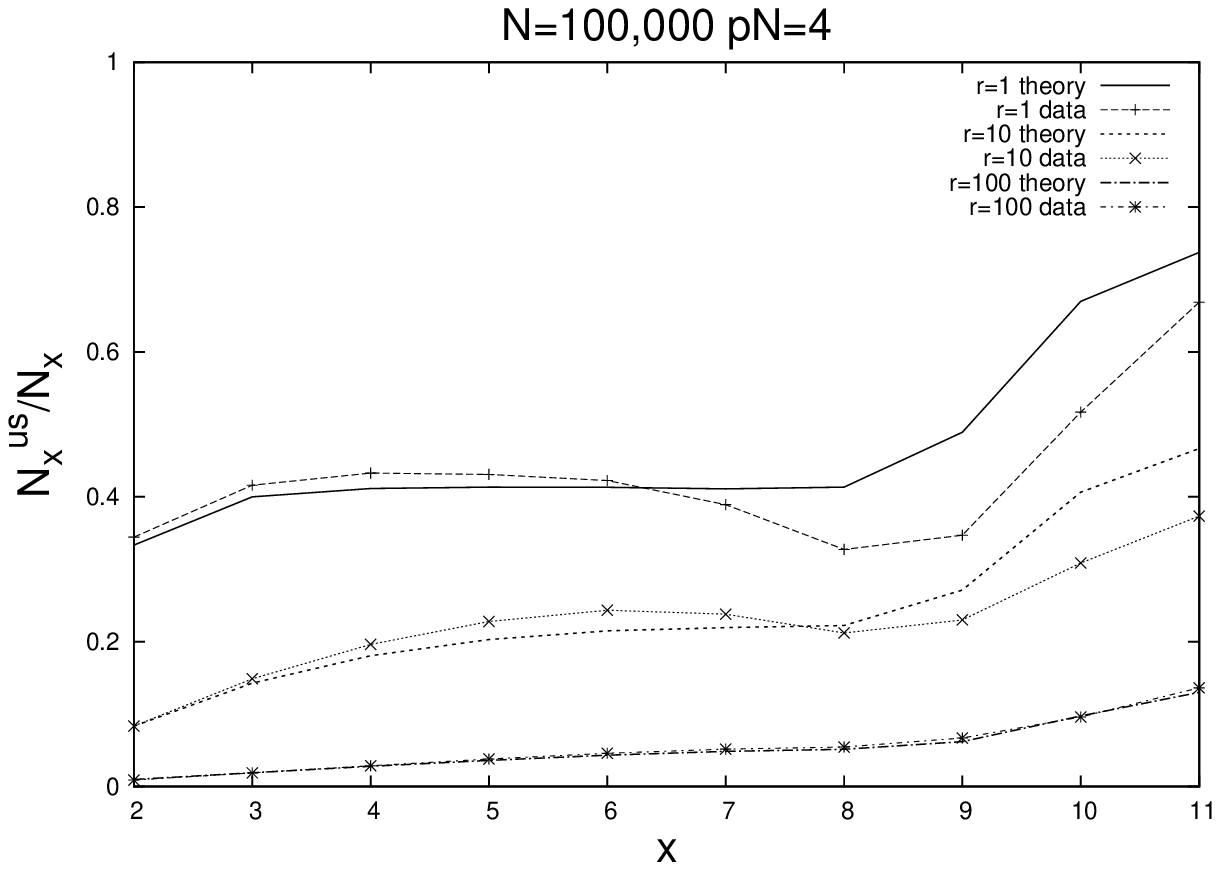}}
\label{fig:nxusN100kd4}
}
\subfigure[]{
  \resizebox{7cm}{!}{\includegraphics[]{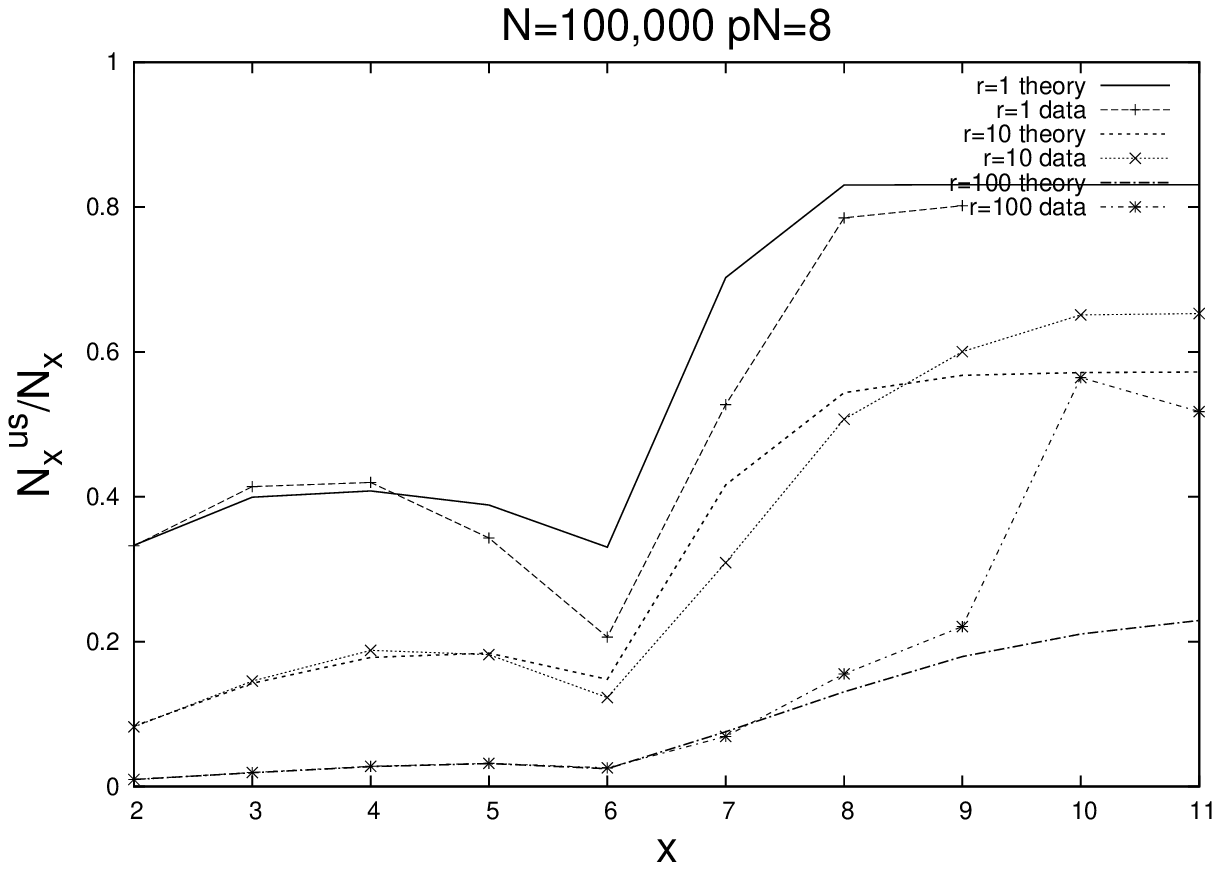}}
\label{fig:nxusN100kd8}
}
\end{center}
\caption{Comparing theory with simulation results: the fraction of unstable nodes at level $x$ : $\Nus_x/N_x$ 
as a function of network size, node degree and $r$.}\label{fig:nxus}
\end{figure}

Figure \ref{fig:nxus} shows in a series of graphs, the numerical evaluation of
equation ~\ref{fig:nxus}, together with simulation results, for network sizes 
network sizes of 1,000 and 100,000, degrees of 4 and 8, and rates of 1, 10 and 
1,000. 

We see that Eq. ~\ref{fig:nxus} predicts simulation results quite well, including the non-monotonic
trend in $\Nus_x/N_x$ as $x$ increases as well as the decrease in  this quantity with increasing $r$.

\graphicspath{{./Final/}}

% NXUS

\end{document}